**Experimental verification of Landauer's principle in erasure of nanomagnetic memory bits**

J. Hong[1], B. Lambson[2], S. Dhuey[3], and J. Bokor[1]

1. Department of Electrical Engineering and Computer Sciences, University of California, Berkeley, California 94720, USA.

2. iRunway, 2906 Stender Way, Santa Clara, CA 95054, USA.

3. The Molecular Foundry, Lawrence Berkeley National Laboratory, 1 Cyclotron Rd., Berkeley, CA 94720, USA.

**In 1961, R. Landauer proposed the principle that logical irreversibility is associated with physical irreversibility and further theorized that the erasure of information is fundamentally a dissipative process.[1] Landauer posited that a fundamental energy cost is incurred by the erasure of information contained in the memory of a computation device.[2] His theory states that to erase one binary bit of information from a physical memory element in contact with a heat bath at a given temperature, at least $k_B T \ln(2)$ of heat must be dissipated from the memory into the environment, where $k_B$ is the Boltzmann constant and $T$ is the temperature. Although this connection between information theory and thermodynamics has proven to be very useful for establishing boundary limits for physical processes, Landauer's principle has been a subject of some debate.[3–8] Despite the theoretical controversy and fundamental importance of Landauer erasure in information technology, this phenomenon has not been experimentally explored using any practical physical implementation for digital information. Here, we report an investigation of the thermodynamic limits of the memory erasure process using nanoscale magnetic memory bits, by far the most ubiquitous digital storage technology**

**today. Through sensitive, temperature dependent magnetometry measurements, we observed that the amount of dissipated energy is consistent (within two standard deviations of experimental uncertainty) with the Landauer limit during an adiabatic erasure process in nanoscale, single domain magnetic thin-film islands. This result confirms the connection between "information thermodynamics" and physical systems and also provides a foundation for the development of practical information processing technologies that approach the fundamental limit of energy dissipation.**

The Landauer erasure process is shown schematically in Fig. 1(A). As shown in Ref. 1, the extracted work from the erasure process is: $W_{Erasure} \geq k_B T \ln(2)$. This energy, $k_B T \ln(2)$, corresponds to a value of 2.8 zJ (2.8×10$^{-21}$ J) at 300K. In the field of ultra-low-energy electronics, computations that approach this energy limit are of considerable practical interest.[9]

Recently, an experimental test of Landauer's principle was reported using a 2-μm-glass bead in water manipulated in a double-well laser trap as a model system.[10] Although the topic is of great importance for information processing in nanodevices, the Landauer limit in single-bit erasure has yet to be tested in potentially practical digital devices.[11] Landauer and Bennett both used nanomagnets as prototypical bistable elements in which the energy efficiency near the fundamental limits was considered.[1,2,12] Currently, magnetic hard-disk storage is ubiquitous, and continues to advance.[13,14] Domain-wall (DW) motion magnetic memory[15], nanomagnetic logic (NML) devices[16], and other types of spintronic devices[17] are very promising new prospects because of their potential for low energy consumption, as well as non-volatility and radiation-hardness. Therefore, it is of great interest to study Landauer's principle in devices using nanomagnets for further energy reductions in the field of future electronics.

The fact that mesoscopic single-domain nanomagnets comprising more than 10$^4$ individual spins can nevertheless behave as a simple system with a single informational degree of freedom has been explicitly analyzed and confirmed theoretically and experimentally.[18,19] Further theoretical studies,[20] in which the adiabatic "reset to one" (erasure) sequence for a nanomagnet memory suggested by Bennett[12] was explicitly simulated using the stochastic Landau-Lifschitz-Gilbert formalism, confirmed Landauer's limit of energy dissipation of $k_B T \ln(2)$ with high accuracy. For a nanomagnetic memory bit, magnetic anisotropy is used to create an "easy axis" along which the net magnetization aligns to minimize magnetostatic energy. The magnetization can align either "up" or "down" along the easy axis to represent binary "0" and "1". We denote the easy axis as the y-axis. The orthogonal x-axis is referred to as the "hard axis". The anisotropy of the magnet creates an energy barrier for the magnetization to align along the hard axis allowing the nanomagnet to retain its state in the presence of thermal noise. To perform Landauer erasure of a bit stored in a nanomagnet, magnetic fields along both the x and y axes are required. The x-axis field is used to lower the energy barrier between the two states, and the y-axis field is then used to drive the nanomagnet into the "1" state.

In the theoretical simulations of Ref. 20, and shown in Fig. 1(B), the Landauer erasure sequence can be divided into four steps. Initially, the nanomagnet is in either "0" or "1" state with equal probability, and afterwards it is reset to the "1" state with unit probability. The internal energy dissipation in the nanomagnet is found by integrating the area of *m-H* loops for both the x- and y-axis and subtracting. To perform the hysteresis loop measurements of interest, the external magnetic fields are specified as a function of time in a quasi-static manner as illustrated in Fig. 1(B). Applying the fields in this manner splits the operation into four stages, and during any given stage, one of the fields is held fixed while the other increased linearly from zero to its maximum value or *vice versa*, as shown in Fig. 1(B). In Stage 1, $H_x$ is applied to saturate the hard axis, which removes the energy barrier, and ensures that the energy dissipation is independent of the barrier height.

Magneto-optic Kerr Effect (MOKE) in the lateral geometry was used to measure the in-plane magnetic moment, *m*, of a large array of identical Permalloy nanomagnets, while the magnetic field, H, was applied using a two-axis vector electromagnet. The experimental set-up is shown in Fig. 2(A). The lateral dimensions of the nanomagnets were less than 100 nm to ensure they were single-domain, while the spacing between magnets was 400 nm, to avoid dipolar interactions between magnets, yet provide sufficient MOKE signal. Scanning electron microscope (SEM) images of the sample are shown in Fig. 2(B). Magnetic force microscopy (MFM) was used to confirm that the nanomagnets have single domain structure and have sufficient anisotropy to retain state at room temperature, as shown in Fig. 2(C). Lateral MOKE probes magnetization along only one in-plane direction,[19] so the sample was mounted on a rotation stage, and separate measurements were made with the sample oriented to measure $m$ along each of the easy and hard axis of the nanomagnets. For each measurement along the two orientations, the magnetic field was slowly (time scale of many seconds) ramped between positive and negative values, according to the Landauer erasure protocol shown in Fig. 1(B). The comprehensive hysteresis loops during the complete erasure process are illustrated schematically in Video S1 in the Supplemental Information.

In order to determine quantitatively the net energy dissipation during the erasure of the memory bit from the MOKE data, it is necessary to calibrate both the applied magnetic field as well as the absolute magnetization of the nanomagnets. The applied field was measured using a three-axis Hall probe sensor. To calibrate the MOKE signal, the total moment, $M_sV_T$ for the full sample was measured using a vibrating sample magnetometer (VSM). $M_s$ is the saturation magnetization for the full sample and $V_T$ is the total volume of magnetic the magnetic layer on the sample. An example of experimental results from one run are shown in Fig. 3. The volume of each nanomagnet, $V$, and the number of nanomagnets on the substrate was carefully measured and calibrated using SEM for the lateral dimensions and count, and atomic force microscopy (AFM) for the thickness. (See Methods and Supplementary Information for details.) In this way, the $M_sV$ value for an individual nanomagnet from the MOKE data could be absolutely determined.

The energy dissipation, (the magnetization energy transferred by the applied magnetic field to a nanomagnet) is determined by the total area of the hysteresis curves. As seen in Fig.

3, and the video S1, the x- and y-hysteresis curves are traversed in opposite directions during the course of the Landauer erasure, so that the total energy of the erasure is found by subtracting the area of the y-hysteresis (easy axis) loop from the area of the x-hysteresis (hard axis) loop. Further details concerning the calculation of energy dissipation are given in the Supplementary Information.

Experimental results for energy dissipation are shown in Fig. 4. The average dissipation for five trials at room temperature was measured to be (6.09 ± 1.43) $zJ$ at T = 300K. This is consistent with a value of (1.45 ± 0.35)$k_BT$, which is extremely close to the Landauer limit of $k_BT$ $ln(2)$ or 0.69 $k_BT$. The quoted error was determined by combining in quadrature the uncertainties in each of the measured variables: nanomagnet area and thickness, magnetic field calibration, magnetic moment, lithographic variation, and the statistical variation among the 5 trials. A separate set of runs was measured at temperatures varying from 300 to 400K with data shown in the SI. As seen in Fig. S1, the hysteresis loops for both axes individually show a clear systematic temperature dependence that is consistent with micromagnetic simulations,[20] but the temperature dependence of the measured net energy dissipation was smaller than the run-to-run variation, as seen in Fig. S2. For this set of runs, the energy dissipation was measured to be (4.2 ± 0.9) $zJ$, which corresponds with a value of (1.0 ± 0.22)$k_BT$ (for T = 300K).

A small remanent magnetization was typically observed in the hard-axis (x) direction (curves in Fig. 3 do not pass through the origin). We attribute this to fabrication variations among the nanomagnets. When the symmetry axes of the individual nanomagnets are not perfectly aligned with the axes of the applied magnetic fields, each of their remanent easy-axis magnetizations will have a small component along the hard axis direction. Due to fabrication variations, there will be a distribution of misalignments. Experiments involving small rotations of the sample to find the net symmetry axis and measure the effect of a net tilt of the array with respect to the magnetic field axes are described in the SI. Based on these experiments, we estimate the magnitude of random variation of the symmetry axes of the nanomagnets across the array to be approximately ±1 degree which is roughly consistent with the observed remanence (see simulation in Fig. S5(C) as well). We suggest that the small excess energy dissipation above the Landauer value that we observe in our experiment is mainly due to this effect. It is discussed in more detail in the SI.

     In conclusion, we have demonstrated for the first time the minimum energy dissipation during an erasure procedure using a digital memory bit on the nanometer scale and confirmed the connection between information thermodynamics and physical systems. Although any practical nanomagnetic memory or logic device will inevitably involve additional energy loss associated with the actuation mechanism (i.e. the external applied magnetic fields in this experiment), these results demonstrate the potential of nanomagnetic devices toward significant reduction of the energy dissipation of future information processing systems.

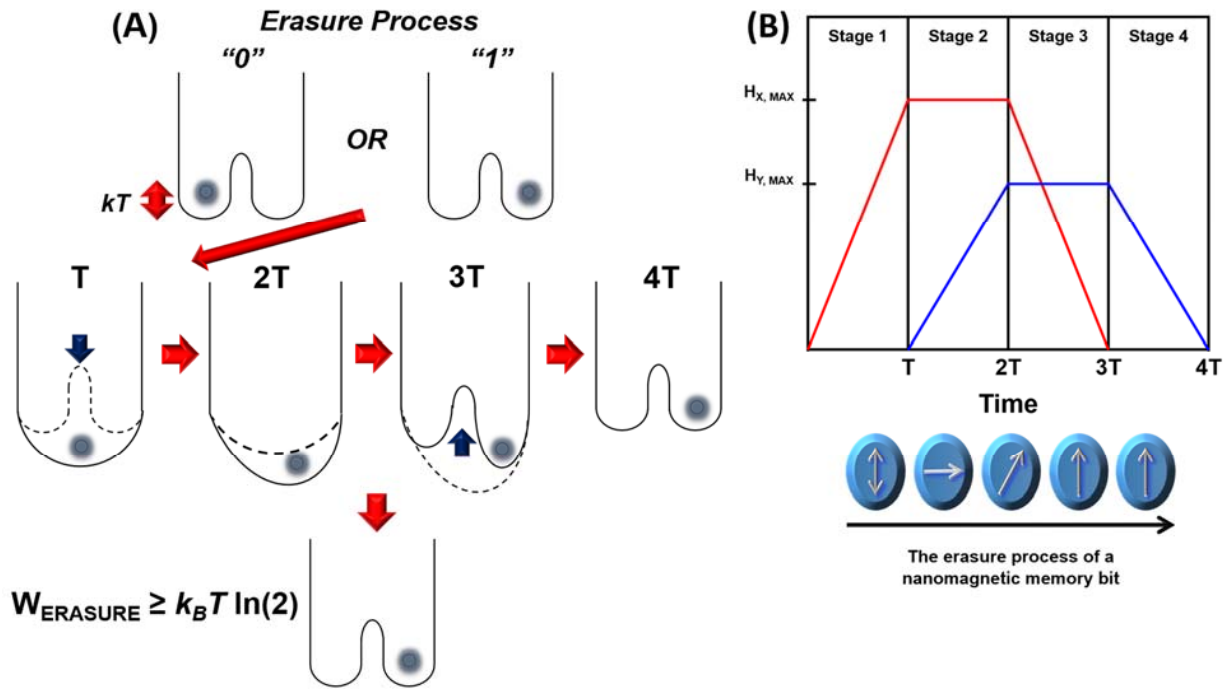

**Figure 1.** │ **Thermodynamics background:** (A) A description of the process of Landauer erasure. Before the erasure, the memory stores information in state "0" or state "1"; after the erasure, the memory stores information in state "0" in accordance with the unit probability. (B) Timing diagram for the external magnetic fields applied during the restore-to-one process. $H_x$ is applied along the magnetic hard axis to remove the uniaxial anisotropy barrier, whereas $H_y$ is applied along the easy axis to force the magnetization into the "1" state. Illustrations are provided of the magnetization of the nanomagnet at the beginning and end of each stage and of the direction of the applied field in the x-y plane.

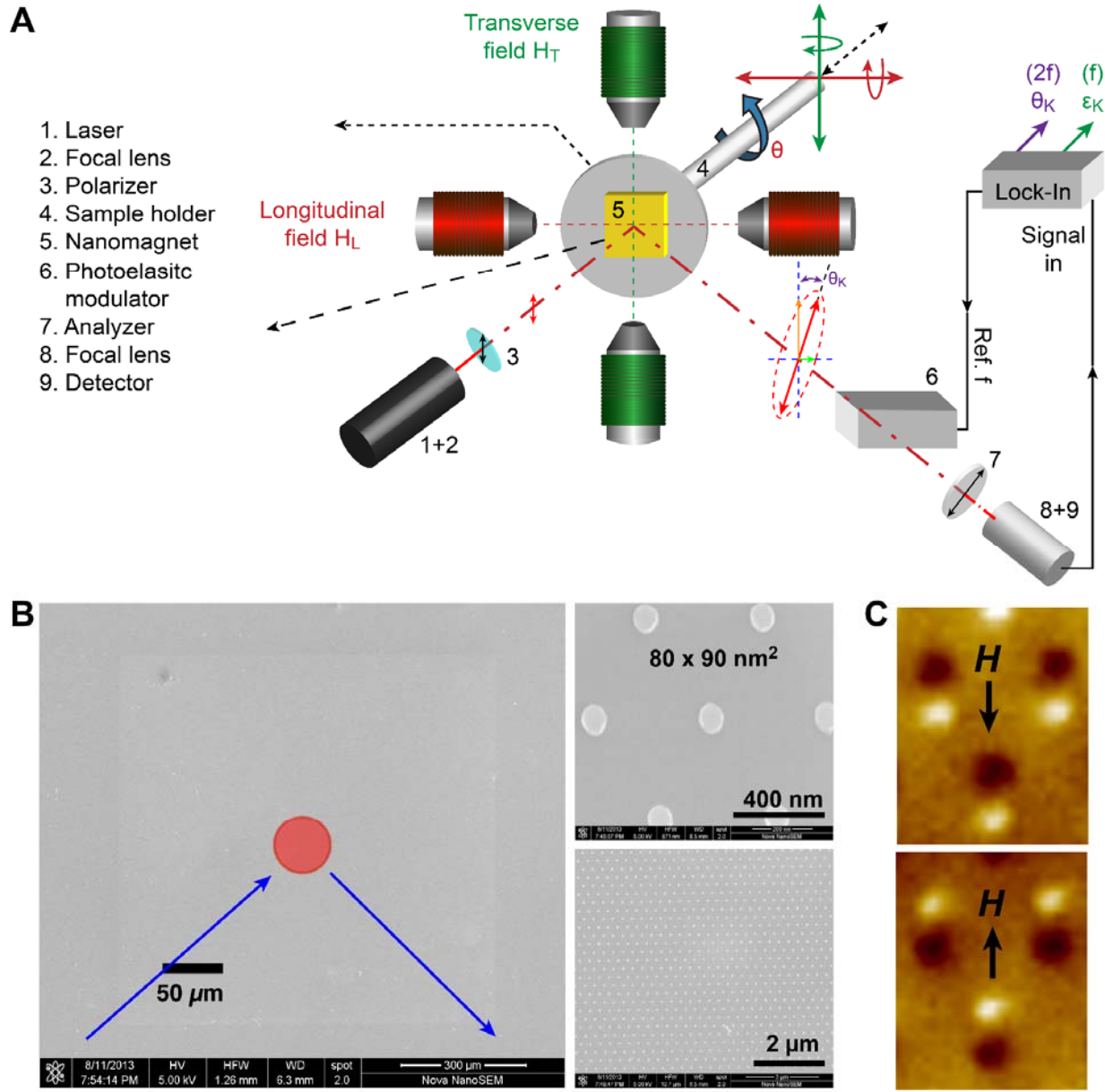

**Figure 2. | The magneto-optical Kerr microscopy experimental set up of Landauer erasure.** (A) Schematic of experimental MOKE set up. (B) Scanning electron microscope (SEM) images of the sample. The circle represents the approximate size of the probe laser spot. (C) Magnetic force microscopy (MFM) images of individual single domain nanomagnets.

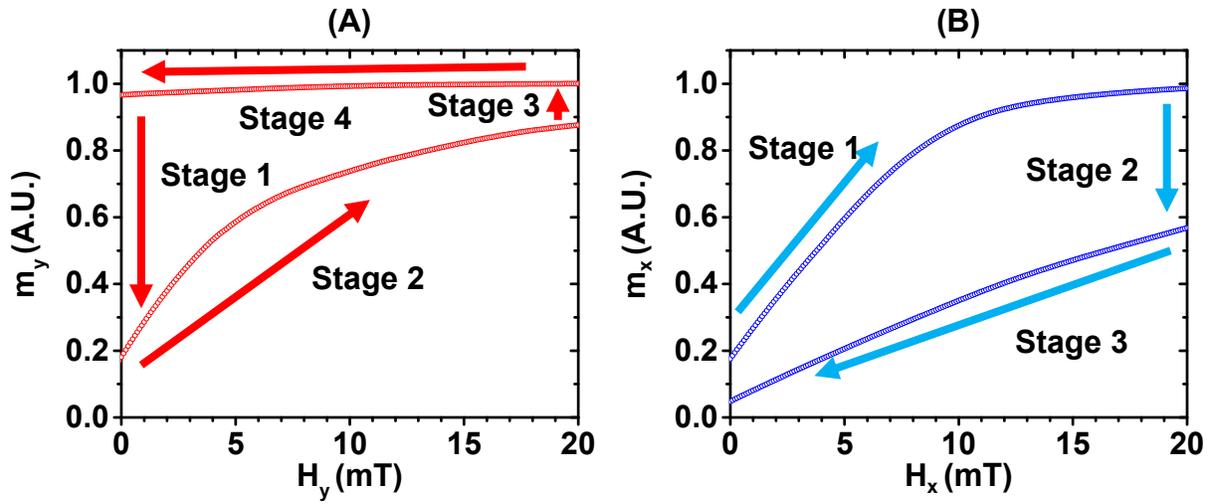

**Figure 3.** | **The experimental m-H hysteresis loops of nanomagnets during Landauer erasure.** (A) The $m_y$-$H_y$ loop (easy axis) and (B) the $m_x$-$H_x$ loop (hard axis). The indicated Stages correspond to the timing diagram shown in Fig. 1(B).

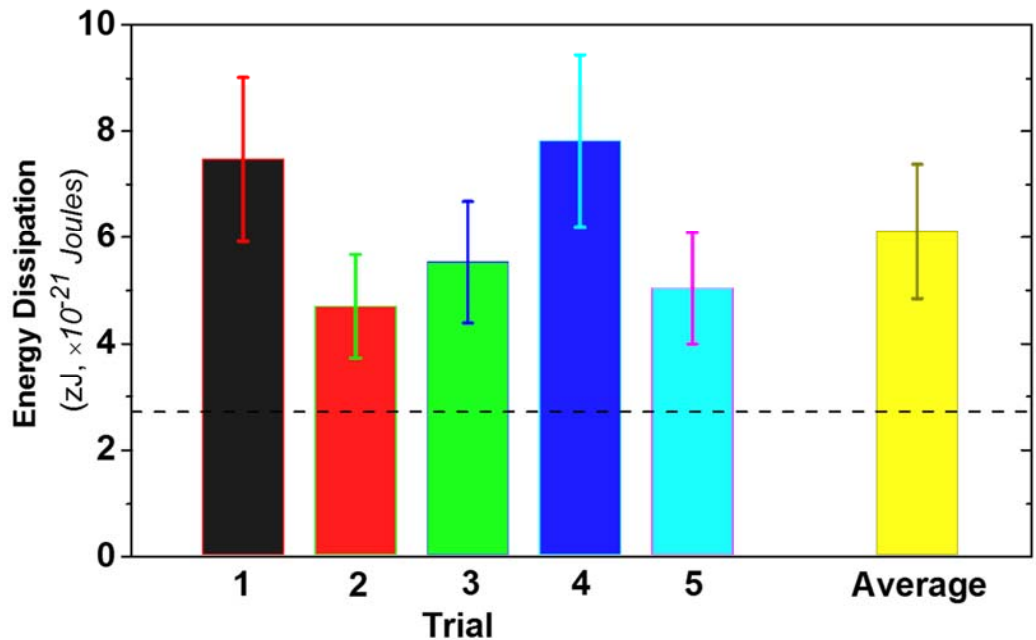

**Figure 4.** |**The experimentally determined energy dissipation during Landauer erasure.** Different bars from 1 to 5 represent experimental attempts to measure energy dissipation. The experimental error was determined from the variables on the table. The dotted line represents the Landauer limit, $k_B T \ln(2)$ for *T=300K*.

**References**

1. Landauer, R. Irreversibility and heat generation in the computing process. *IBM J. Res. Develop.* **5**, 183−191 (1961).
2. Landauer, R. Dissipation and noise immunity in computation and communication. *Nature* **335**, 779−784 (1988).
3. Porod, W., Grondin, R. O., Ferry, D. K., Porod, G. Dissipation in computation. *Phys. Rev. Lett.* **52**, 232−235 (1984).

4. Benioff, P. Comment on "Dissipation in Computation." *Phys. Rev. Lett.* **53**, 1203–1203 (1984).

5. Bennett, C. H. Thermodynamically reversible computation. *Phys. Rev. Lett.* **53**, 1202–1202 (1984).

6. Landauer, R. Dissipation in computation. *Phys. Rev. Lett.* **53**, 1205–1205 (1984).

7. Toffoli, T. Comment on "Dissipation in Computation." *Phys. Rev. Lett.* **53**, 1204–1204 (1984).

8. Porod, W., Grondin, R. O., Ferry, D. K., Porod, G. Porod et al. Respond. *Phys. Rev. Lett.* **53**, 1206–1206 (1984).

9. Meindl, J. D., Davis, J. A. The fundamental limit on binary switching energy for terascale integration. *IEEE J. Solid-state Circuits* **35**, 1515–1516 (2000).

10. Berut, A., Arakelyan, A., Petrosyan, A., Ciliberto, S., Dillenshneider, R., Lutz, E. Experimental verification of Landauer's principle linking information and thermodynamics. *Nature* **484**, 187–190 (2012).

11. Sagawa, T. Thermodynamics of information processing in small systems. *Prog. Theor. Phys.* **127**, 1–56 (2012).

12. Bennett, C. H. The thermodynamics of computation - a review. *Int. J. Theor. Phys.* **21**, 905–940 (1982).

13. Thompson, D. A., Best, J. S. The future of magnetic data storage technology. *IBM J. Res. Develop.* **44**, 311–321 (2000).

14. Andreas, M., Kentaro, T., David, T. M., Manfred, A., Yoshiaki, S., Yoshihiro, I., Shouheng, S., Fullerton, E. F. Magnetic recording: advancing into the future. *J. of Phys. D: Appl. Phys.* **35**, R157 (2002).

15. Parkin, S. S. P., Hayashi, M., Thomas, L. Magnetic domain-wall racetrack memory. *Science* **320**, 190 (2008).

16. Imre, A., Csaba, G., Ji, L., Orlov, A., Bernstein, G. H., Porod, W. Majority logic gate for magnetic quantum-dot cellular automata. *Science* **311**, 205−208 (2006).

17. Wolf, S. A., Chtchelkanova, A. Y., Treger, D. M. Spintronics- A retrospective and perspective. *IBM J. Res. Develop.* **50**, 101−110 (2006).

18. Salahuddin, S., Datta, S. Interacting systems for self-correcting low power switching. *App. Phys. Lett.* **90**,093503 (2007).

19. Allwood, D. A., Xiong, G., Cooke, M. D., Cowburn, R. P. Magneto-optical Kerr effect analysis of magnetic nanostructures. *J. Phys. D: Appl. Phys.* **36**, 2175−2182 (2003).

20. Lambson, B., Carlton, D., Bokor, J. Exploring the thermodynamic limits of computation in integrated systems: magnetic memory, nanomagnetic logic, and the Landauer limit. *Phys. Rev. Lett.* **107**, 010604 (2011); Madami, M., d'Aquino, M. , Gubbiotti, G., Tacchi, S., Serpico, C., and Carlotti, G. "Micromagnetic study of minimum-energy dissipation during Landauer erasure of either isolated or coupled nanomagnetic switches," *Physical Review B,* **90**, 104405, (2014).
**METHODS SUMMARY**

***Fabrication of Nanomagnetic Memory Bits.*** We fabricated an array of identical, single-domain, non-interacting, elliptically shaped nanomagnets by lift-off patterning of an e-beam evaporated amorphous Permalloy (NiFe) film on a silicon substrate using electron-beam (e-beam) lithography. The sample was fabricated at the Molecular Foundry at Lawrence Berkeley National Laboratory.

***Dimensional Metrology of Nanomagnet Islands using Scanning Electron Microscopy and Scanning Probe Microscopy.*** SEM images were collected with a Carl Zeiss LEO 1550. The statistical errors of area of the magnet and thickness measurements are 1.2 % and 7.9%, respectively. The image analysis was performed using *ImageJ* software from National Institute of Health (NIH, Bethesda, MD). The average size of magnets was calculated with the software, and calibrated to the highly accurate average pitch of the nanomagnet array produced by the e-beam lithography tool (Vistec VB300). The thickness of the nanomagnets was determined using Atomic Force Microscopy (AFM) performed in non-contact mode using a Veeco Dimension 3100 system. MFM measurements using the same instrument were conducted in a dynamic lift mode with a lift-off distance of 30 nm.

***Magneto-Optical Kerr Spectroscopy.*** To perform high-resolution magneto-optical Kerr spectroscopy, we used a focused MOKE system in lateral mode. A 635 nm diode laser was directed toward the sample, which was located between the poles of a vector magnet. The laser spot size on the sample was approximately 50 $\mu$m, covering approximately $10^4$ nanomagnets. The magnetic field at the probe spot was calibrated by a three-axis Hall probe sensor (C-H3A-2m Three Axis Magnetic Field Transducer, SENIS GmbH Zürich, Switzerland). The accuracy of the magnetic field measurement is estimated at ~1%. The time to sweep full hysteresis loops was 20 minutes (1 Oe/s).

***Vibrating Sample Magnetometry***. The saturation magnetization measurements were performed using a VSM 7400 of Lake Shore Cryotronics with a 3.1-T electromagnet (See Fig. S1). The sample was mounted on a quartz holder using vacuum grease. The magnetic moment was measured and averaged between upper and lower values of the curves on both easy axis and hard axis. Further details are provided in the Supplementary Information.The average magnetic moment for the full sample is 9.70 x $10^{-9}$ $Am^2$ with 10% uncertainty.

**Acknowledgements**

We gratefully acknowledge the financial support from the NSF Center for Energy Efficient Electronics Science under Award number 0939514. In addition, the authors would like to thank Professor Takahiro Sagawa from the University of Tokyo for the valuable and thoughtful discussions we had concerning this study. The work at the Molecular Foundry was supported by

**Author Information**

Department of Electrical Engineering and Computer Sciences, University of California, Berkeley, CA 94720 USA.

# Supplemental Information File

**Experimental verification of Landauer's principle in erasure of nanomagnetic memory bits**

J. Hong, B. Lambson, S. Dhuey, and J. Bokor

Temperature dependence of m-H loops and energy dissipation

The sample was mounted on a heater stage, and the measurement was repeated for a range of temperatures between 300 and 400K. Figure S1 shows the hysteresis loops for both axes. A clear temperature dependence is shown from the measurement. Experimental results for energy dissipation at different temperatures are shown in Fig. S2. The average dissipation for this set of runs was measured to be (4.2 ± 0.9) zJ, or (1.0 ± 0.22)$k_BT$, (for T = 300K) which is consistent with the results of the 5 runs that were measured at 300K shown in Fig. 4.

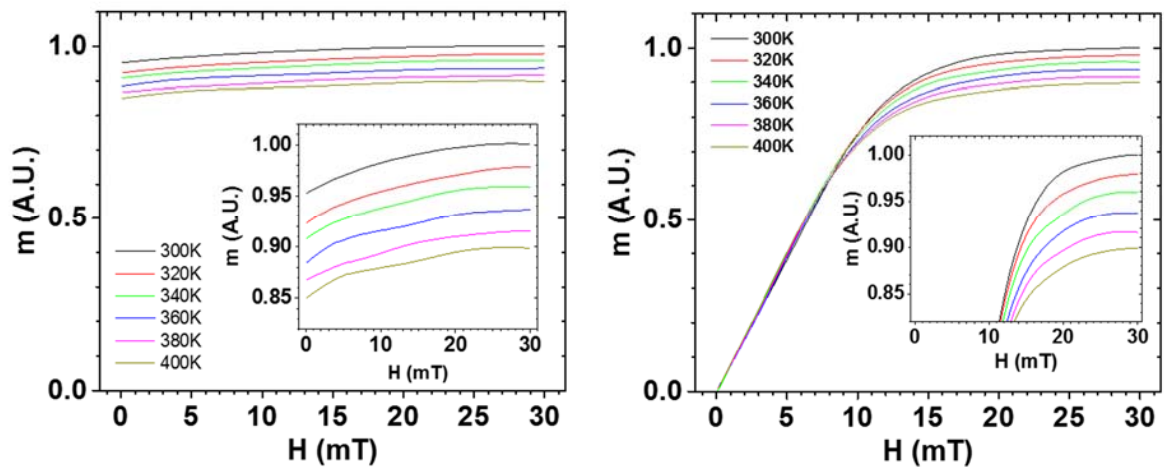

**Figure S1.** | **The temperature dependence of energy dissipation during Landauer erasure**. Triangles represent experimental data obtained from integrating and subtracting hysteresis loops similar to the example shown in Fig. 3. The red line is the best fit to the experimental data. The black squares represent the Landauer limit, $k_B T \ln(2)$.

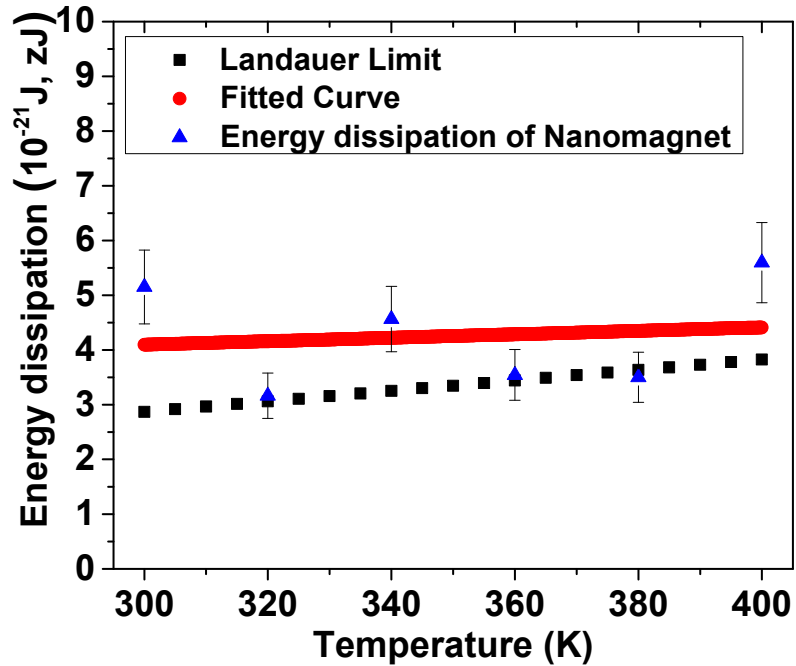

**Figure S2.** | The temperature dependence of the measured energy dissipation.

Vibrating Sample Magnetometry (VSM) and the estimation of the collective spin moment, *μ*

The total magnetic moment of the patterned nanomagnetic bits was measured by vibrating sample magnetometry, as shown in Fig. S3. The measured magnetic moment of the structure is $9.70 \times 10^{-6}$ emu. The measured average volume of the magnets is $2.51 \times 10^{-25}$ m$^3$ with 8 % uncertainty. Based on these dimensions, we calculate magnetization of $M_s = 7.91 \times 10^5$ A/m and the individual spin moment values, *μ* (= $M_S V$) of each individual magnet was calculated as $2.00 \times 10^{-16}$ Am$^2$.

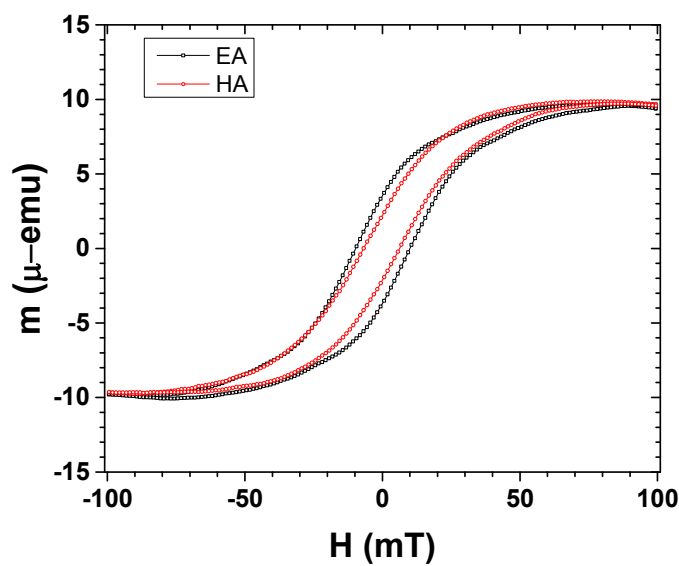

**Figure S3.** | m-H loops of the total magnetic moment of the full sample. The magnetic moment values of the upper and lower hysteresis loops were measured and averaged for the total magnetization to estimate the single spin moment $M_SV$.

Measured m-H loops of the full sample along both the easy axis and hard axis are shown in Fig. S3. On this sample, 9 different arrays of different size of nanomagnets were fabricated in different areas on the same substrate. This is the reason that the m-H loops measured by VSM show much larger remanence along the hard axis direction than the MOKE measurements on a small area of the sample with nominally uniform nanomagnet size and shape. The main objective for this experiment was to measure the saturation magnetization value for the magnetic material in order to calibrate the saturated magnetization level in the MOKE data.

Lithographic Variations

Nanomagnet arrays patterned using e-beam lithography show dimensional variations.[21, 22] We believe that the main effects of these variations on this experiment are (1) deviations of the symmetry axis of individual nanomagnets from the net symmetry axis of the nanomagnet array (the tilt effect, illustrated in Fig. S3B), and (2) resulting dependence of the dissipation on $H_{x,\,max}$.

We studied the tilt effect in both experiment and simulation and considered the contribution stemming from the lithographic tilt averaged over the measurement cycles corresponding to changes of state as well as over the cycles in which the memory remained unchanged.

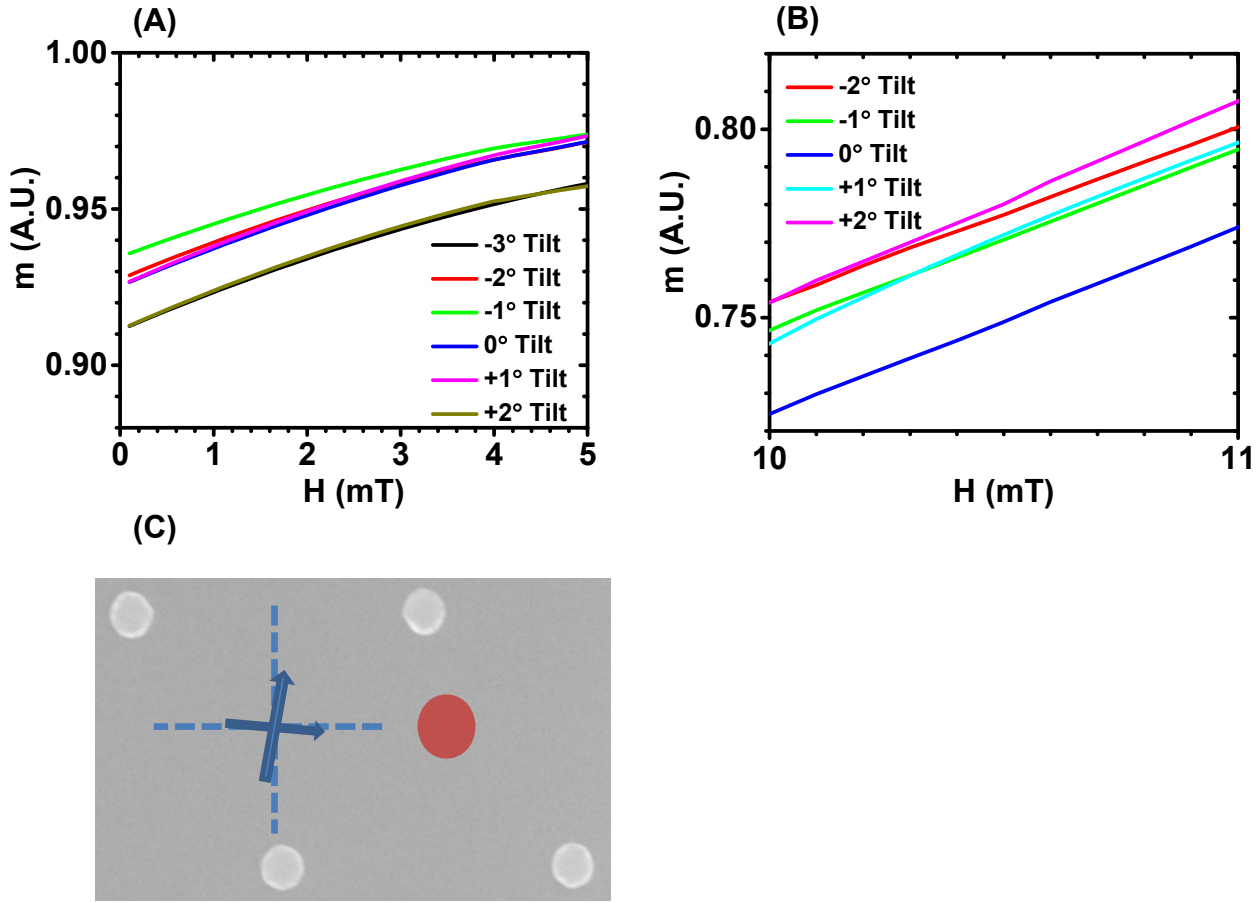

**Figure S4.** | (A) Hard axis *m-H* curves corresponding to stage 1 of the Landauer erasure protocol with various sample tilt angles. (B) m-H loops of easy axis in stage 4 with sample tilt angles. (C) A SEM image with the schematics of experimentally determined easy and hard axes of the nanomagnet.

To gain insight into the tilt effect, we rotated the sample by small amounts around the net symmetry direction and repeated the measurement. Fig. S4(A) shows m-H loops with different tilt angles in stage1. Stage 4 easy axis curves as a function of sample tilt (expanded to show the variation) are shown in Fig. S4(B). From the experiment, we observed lithographic tilt effect in both the easy axis and hard axis. The schematic scenario of the effect in a SEM image is shown in Fig. S4(C). Due to lithographic variations, each nanomagnet has a small random effective tilt with respect to the average symmetry axis.

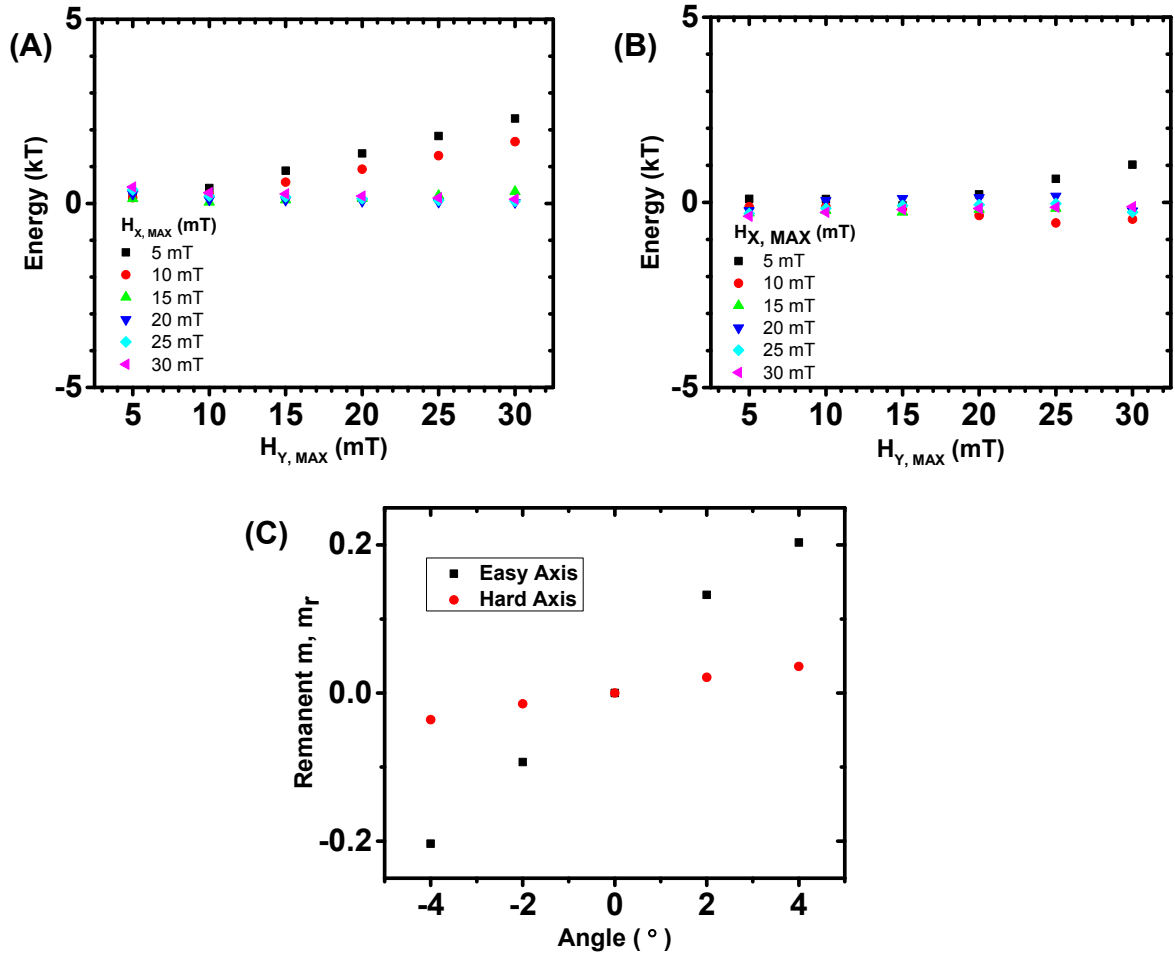

**Figure S5.** | (A) The simulated energy dissipation at 0 K by varying the maximum fields ($H_{x,max}$ and $H_{y,max}$) field. (B) The energy dissipation calculation from simulation with tilt effect (2 degrees tilt and average) by increasing $H_{x,max}$ and $H_{y,max}$. (C) Simulated remanent magnetic moment with different tilt angles in hard axis and easy axis corresponding to stages 2 and 3, respectively.

With small tilt (±2 degrees), a noticeable increase in energy dissipation was observed. In Fig. S5(A) and S5(B), we show the results of micromagnetc simulations of the effects of small tilt angle combined with variation of the perpendicular field values $H_{x,max}$ and $H_{y,max}$. For these simulations, no thermal fluctuations were included, so that the ideal value of energy dissipation should be zero. The remanent magnetic moments with different tilt angles are shown in Fig. S5(C). As seen in Fig. S5(A, B), for small values of $H_{x,max}$ (hard axis field), the excess dissipation increases with increasing $H_{y,max}$ because the magnet is not fully nulled. These results indicate that with $H_{x,max}$ and $H_{y,max}$ in the 10 ~ 15 mT range, the excess energy dissipation for small tilt angles is approximately **±0.25 $k_B$T (T = 300K)**, which is the dominant term in our quoted experimental uncertainty and is indicated as "Lithography" in the table of error terms in Fig. 4. Based on these results from both the simulation and experimental study of tilt effects, we chose 15 mT for $H_{x,max}$ and 10 mT for $H_{y,max}$ in our experimental runs.

*2. Online Video*

The temporal sequence of the application of magnetic fields in the experiment is illustrated in this video.

**References**

21. Lee, B., Hong, J., Amos, N., Dumer, I., Litvinov, D., Khizroev, S. Sub-10-nm-resolution electron-beam lithography toward very-high-density multilevel 3D nano-magnetic information devices. *J. Nanopart. Res.* **15**, 1-8 (2013).

22. Gu, Z., Nowakowski, M. E., Carlton, D. B., Storz, R., Hong, J., Chao, W., Lambson, B., Bennett, P., Alam, M. T., Marcus, M. A., Doran, A., Young, A., Scholl, A., and Bokor, J. Speed and Reliability of Nanomagnetic Logic Technology. arXiv:1403.6490. (2014).